\documentclass[aip,amsmath,amssymb,reprint]{revtex4-1}

\usepackage{graphicx}
\usepackage{dcolumn}
\usepackage{bm}

\usepackage{subfigure}
\usepackage{booktabs}
\usepackage{lineno}
\usepackage{mathptmx}

\usepackage[utf8]{inputenc} 
\usepackage[T1]{fontenc}    
\usepackage{hyperref}       
\usepackage{url}            
\usepackage{booktabs}       
\usepackage{amsfonts}       
\usepackage{nicefrac}       
\usepackage{microtype}      
\usepackage{lmodern}      

\usepackage{amsmath}
\usepackage{amssymb}

\usepackage{float}

\graphicspath{{./},{./doc},{./figures/},{./figures/executions/Baseline/},{./figures/linear/},{./figures/nonlinear/},{./figures/movies/f697bb/}}

\usepackage{tikz}
\usetikzlibrary{arrows,positioning}

\usepackage{xcolor}

\newcommand{\eqnref}[1]{Eqn. (\ref{eqn:#1})}
\newcommand{\figref}[1]{Fig. \ref{fig:#1}}
\newcommand{\secref}[1]{\S\ref{sec:#1}}

\renewcommand{\vec}[1]{{\mathbf{#1}}}

\newcommand{\weighting}{c}

\definecolor{darkgreen}{RGB}{0,128,0}
\newcommand{\macrocolor}{black}

\newcommand{\transformLetter}{\textcolor{\macrocolor}{\vec \theta}}
\newcommand{\trans}{\transformLetter}
\newcommand{\transInv}{\transformLetter^{-1}}
\newcommand{\transLearned}{\hat{\transformLetter}}
\newcommand{\transInvLearned}{\hat{\transformLetter}^{-1}}
\newcommand{\symplecto}{\textcolor{\macrocolor}{S}}

\newcommand{\pder}[2]{\textcolor{\macrocolor}{\frac{\partial #1}{\partial #2}}}
\newcommand{\diff}{\textcolor{\macrocolor}{\mathrm{d}}}
\newcommand{\Hder}[1]{\pder{\hat{H}}{#1}}

\newcommand{\subfigwidthfrac}{0.45}

\newcommand{\imagex}{\vec x}

\newcommand{\hatzvec}{\textcolor{\macrocolor}{\hat{\vec{z}}}}
\newcommand{\hatqvec}{\textcolor{\macrocolor}{\hat{\vec{q}}}}
\newcommand{\hatpvec}{\textcolor{\macrocolor}{\hat{\vec{p}}}}

\begin{document}

\author{Tom Bertalan}
\affiliation{Department of Mechanical Engineering, The Massachusetts Institute of Technology, Cambridge, Massachusetts 02139, USA}
\author{Felix Dietrich}
\affiliation{Department of Applied Mathematics and Statistics, Johns Hopkins University, Baltimore, Maryland 21211, USA}
\author{Igor Mezi\'{c}}
\affiliation{Department of Mechanical Engineering, The University of California Santa Barbara, Santa Barbara, California 93106, USA}
\author{Ioannis G. Kevrekidis}
\affiliation{Department of Chemical and Biomolecular Engineering, Johns Hopkins University, Baltimore, Maryland 21211, USA}
\email{yannisk@jhu.edu}

\title{On Learning Hamiltonian Systems from Data}

\begin{abstract}
Concise, accurate descriptions of physical systems through their conserved quantities abound in the natural sciences. In data science, however, current research often focuses on regression problems, without routinely incorporating additional assumptions about the system that generated the data.
Here, we propose to explore a particular type of underlying structure in the data: Hamiltonian systems, where an ``energy''  is conserved.
Given a collection of observations of such a Hamiltonian system over time, we extract phase space coordinates and a Hamiltonian function of them that acts as the generator of the system dynamics.
The approach employs an autoencoder neural network component to estimate the transformation from observations to the phase space of a Hamiltonian system. An additional neural network component is used to approximate the Hamiltonian function on this constructed space, and the two components are trained jointly. As an alternative approach, we also demonstrate the use of Gaussian processes for the estimation of such a Hamiltonian.
After two illustrative examples, we extract an underlying phase space as well as the generating Hamiltonian from a collection of movies of a pendulum. The approach is fully data-driven, and  does not assume a particular form of the Hamiltonian function.
\end{abstract}

\keywords{Hamiltonian systems, neural networks, Gaussian processes}

\maketitle

\begin{quotation}
Neural network-based methods for modeling dynamical systems are becoming again widely used, and methods that explicitly learn the physical laws underlying continuous observations in time constitute a growing subfield. Our work contributes to this thread of research by incorporating additional information into the learned model; namely, the knowledge that the data arise as observations of an underlying Hamiltonian
system.

We use machine learning to extract models of systems whose dynamics conserve a particular quantity (the Hamiltonian).
We train several neural networks to approximate the total energy function for a pendulum, in both its natural action-angle form and also as seen through several distorting observation functions of increasing complexity. A key component of the approach is the use of automatic differentiation of the neural network in formulating the loss function that is minimized during training.

Our method requires data evaluating the first and second time derivatives of observations across the regions of interest in state space or, alternatively, sufficient information (such as a sequence of delayed measurements) to estimate these. We include examples in which the observation function is nonlinear, and high-dimensional.
\end{quotation}

\section{Introduction}
Current data science exploration of dynamics often focuses on regression or classification problems, without routinely incorporating additional assumptions about the nature of the system that generated the data.
This has started to change recently, with approaches to extract generic dynamical systems by~\cite{schmidt-2009}, specifying the variables and possible expressions for the formulas beforehand. 
%
In particular, treating the central object to be modeled as the discrete-time flowmap 
    $
        \Phi_\tau(\vec x) = 
        \vec x+\int_{0}^\tau \vec f(\vec x(t))
        \diff t
    $
    but learning $\Phi$ directly as a black box
    may result in qualitative differences from the true system\cite{Rico-Martinez1992a}. Using the associated differential equation $\diff \vec x /\diff t=\vec f(\vec x)$ instead,
    allows us to exploit established numerical integration schemes to help approximate the flow map, and can be accomplished with a neural network
    by structuring the loss function
    similarly to such classical numerical integration schemes.
    In addition to our older work along these lines,
        \cite{Gonzalez-Garcia1998b, Rico-Martinez1994, Rico-Martinez1994b, Rico-Martinez1993, Rico-Martinez1995, Rico-Martinez1992a,rico-martinez_noninvertibility_1993,Rico-Martinez2000d},
    more recent work has revived this approach,
    focused on treating the layers of a deep neural network
    as iterates of a dynamical system,
    where ``learning'' consists of discovering the right attractors~\cite{e_proposal_2017,chang_multi-level_2018,lu_beyond_2018}.
    In particular, interest has focused on how
    residual networks\cite{he_deep_2016}
    and highway networks\cite{srivastava_highway_2015}
    can be interpreted 
    as iterative solvers\cite{greff_highway_2017}
    or
    as iterated dynamical systems\cite{chen_neural_2018}.
    In the latter paper (a NeurIPS 2018 best paper awardee),
    the authors
    choose not to unroll the iteration in time explicitly,
    but instead use continuous-time numerical integration.
    While the focus was on the dynamics-of-layers concept,
    time series learning was also performed.

The Koopman operator has also been employed in combination with neural networks to extract conservation laws and special group structures~\cite{kaiser-2018,lusch_deep_2018}.
Symmetries in relation to conserved quantities
are a well-studied problem in physics~\cite{hamilton_general_1834,noether-1971,livio-2012,kondor-2018b}.
A recent thread of research consists of learning physics models from observation data~\cite{mottaghi_newtonian_2016},
including modeling discrete-time data as observations of a continuous-time dynamical system~\cite{raissi_inferring_2017,raissi_hidden_2018}.

%
%

It is informative to study physical systems 
through their conserved quantities,
such as total energy, which can be encoded in a Hamiltonian function.~\cite{hamilton_general_1834,almeida1992}
The measure-preserving property of Hamiltonian systems
has recently been exploited for transport of densities
in Markov-chain Monte-Carlo methods~\cite{brooks_mcmc_2011},
and variational autoencoders~\cite{caterini_hamiltonian_2018,rezende_variational_2015}.
For our purposes, a natural progression of the thread of computational modeling of physics from observations
is to represent the Hamiltonian function directly.

Concurrently with this submission,
    two papers independently addressing similar issues appeared as  preprints\cite{greydanus_hamiltonian_2019,toth_hamiltonian_2019}. In the first of these\cite{greydanus_hamiltonian_2019}, a loss function very similar to parts of our \eqnref{untransformedloss} was used. The second paper focuses on the transformation of densities generated through the Hamiltonian.
    It is possible to draw some analogies between the second preprint and the old (non-Hamiltonian) work we mentioned above\cite{Gonzalez-Garcia1998b, Rico-Martinez1994, Rico-Martinez1994b, Rico-Martinez1993, Rico-Martinez1995, Rico-Martinez1992a,rico-martinez_noninvertibility_1993,Rico-Martinez2000d},
    as this newer work also used rollout with a time step templated 
    on a classical numerical integration method (here, symplectic Euler and leapfrog).
    Both papers also use the pendulum as an example, emphasizing conditions where the system can be well approximated by a linear system: a thin annulus around the stable steady state at $(q,p)=(0,0)$ where the trajectories are nearly circular.

The remainder of the paper is structured as follows:
\begin{enumerate}
\item We derive data-driven approximations (through two approaches: Gaussian processes and neural networks) of a Hamiltonian function on a given phase space, from time series data.
The Hamiltonian functions we consider do not need to be separable as a sum $H(q,p)=T(p)+V(q)$, and in our illustrative example we always work in the fully nonlinear regime of the pendulum.
\item We build data-driven reconstructions of a phase space from (a) linear and (b) nonlinear, non-symplectic transformations of the original Hamiltonian phase space. The reconstruction then leads to a symplectomorphic copy of the original Hamiltonian system.
\item We construct a completely data-driven pipeline combining (a) the construction of an appropriate phase space and (b) the approximation of a Hamiltonian function on this new phase space, from nonlinear, high-dimensional observations (e.g. from movies/sequences of movie snapshots).
\end{enumerate}

\section{General description}
A Hamiltonian system on Euclidean space $E=\mathbb{R}^{2n}$, $n\in\mathbb{N}$ is determined through a function $H:E\to\mathbb{R}$ that defines the equations
\begin{eqnarray}
\dot q(t)&=&\partial H(q(t),p(t)) / \partial p,\label{eqn:hamiltonian system 1}\\
\dot p(t)&=&-\partial H(q(t),p(t)) / \partial q,\label{eqn:hamiltonian system 2}
\end{eqnarray}
where $\dot{(\quad)}:=\diff/\diff t$, and $q(t),p(t)\in\mathbb{R}^n$ are interpreted as ``position'' and ``momentum'' coordinates in the ``phase space'' $E$. In many mechanical systems, and in all examples we discuss in this paper, the interpretation of the coordinates $q,p$ is reflected in the dynamics through $\dot q=p$, i.e. $H(q,p)=\frac{1}{2}p^2+h(q)$ for some function $h:\mathbb{R}^n\to\mathbb{R}^n$. In general, the equations (\ref{eqn:hamiltonian system 1}, \ref{eqn:hamiltonian system 2}) imply that the Hamiltonian is constant along trajectories $(q(t),p(t))$, because
\begin{equation}
\label{eqn:Hdot}
\frac{\diff}{\diff t}H(q,p)=\frac{\partial H}{\partial q}(q,p) \cdot \dot {q} + \frac{\partial H}{\partial p}(q,p) \cdot \dot {p}=0.
\end{equation}
Equations (\ref{eqn:hamiltonian system 1}, \ref{eqn:hamiltonian system 2}) can be restated as a partial differential equation for $H$ at every $(q,p)\in E$:
\begin{equation}\label{eqn:H pde}
 \underbrace{\left[\begin{matrix}
0&I\\-I&0
\end{matrix}\right]}_{\omega}\cdot \nabla H(q,p) - \nu(q,p)=0,
\end{equation}
where $I\in\mathbb{R}^{n\times n}$ is the identity matrix and $\nu$ is the vector field on $E$ (the left hand side of (\ref{eqn:hamiltonian system 1}, \ref{eqn:hamiltonian system 2})), which only depends on the state $(q,p)$.  The symplectic form on the given Euclidean space takes the form of the matrix $\omega$.

In the next section, we discuss how to approximate the function $H$ from given data points $D=\{(q_i,\dot q_i, \ddot q_i)\}_{i=1}^N$.
This involves solving the partial differential \eqnref{H pde} for $H$. Since these equations determine $H$ only up to an additive constant, we assume that we also know the value $H_0=H(q_0,p_0)$ of $H$ at a single point $(q_0,p_0)$ in phase space. This is not a major restriction for the approach, because $H_0$ as well as $(q_0,p_0)$ can be chosen arbitrarily.

\section{Example: the nonlinear pendulum}
As an example, consider the case $n=1$, and the Hamiltonian
\begin{equation}
\label{eqn:H true}
H(q,p)=\frac{p^2}{2}+(1-\cos(q)).
\end{equation}
This Hamiltonian forms the basis for the differential equations of the nonlinear pendulum,
$\ddot q=-\sin(q)$,
or, in first-order form,
$\dot q = \partial H(q,p) / \partial p=p$
and
$\dot p = - \partial H(q,p)/\partial q = -\sin(q)$.
%
In this section, we numerically solve PDE (\ref{eqn:H pde}) 
by approximating the solution $H$ using two approaches: Gaussian Processes~\cite{rasmussen-2005} (\secref{gps for h}) and neural networks (\secref{nets for h}).

\subsection{Approximation using Gaussian processes}\label{sec:gps for h}
We model the solution $H$ as a Gaussian Process $\hat{H}$ with a Gaussian covariance kernel,
\begin{equation}
k(x,x')=\exp\left(-\|x-x'\|^2/\epsilon^2\right),
\end{equation}
where $x$ and $x'$ are points in the phase space, i.e. $x=(q,p)$, $x'=(q',p')$ and $\epsilon\in\mathbb{R}^+$ is the kernel bandwidth parameter (we chose $\epsilon=2$ in this paper). Given a collection $X$ of $N$ points in the phase space, as well as the function values $H(X)$ at all points in $X$, the conditional expectation of the Gaussian Process $\hat{H}$ at a new point $y$ is
\begin{equation}\label{eqn:mean of gaussian process}
E[\hat{H}(y)|X,H(X)]=k(y,X)^Tk(X,X')^{-1}H(X),
\end{equation}
where we write $\left[k(X,X')\right]_{i,j}:=k(x_i,x_j)$ for the kernel matrix evaluated over all $x$ values in the given data set $X$.
In \eqnref{mean of gaussian process}, the dimensions of the symbols are $y\in\mathbb{R}^{2n}$, $k(X,X')\in\mathbb{R}^{N\times N}$, $k(y,X)\in\mathbb{R}^{N}$, and $H(X):=(H(x_1),H(x_2),\dots,H(x_N))\in\mathbb{R}^{N}$. All vectors are column vectors.
 Estimates of the solution $H$ to the PDE at new points depend on the value of $H$ over the entire data set. We do not know the values of $H$, but differentiating \eqnref{mean of gaussian process} allows us to set up a system of equations to estimate $H$ at an arbitrary number $M$ of new points $y$ close to the points $x\in X$, by using the information about the derivatives of $H$ given by the time derivatives $\dot q$ in \eqnref{hamiltonian system 1} and (the negative of) $\dot p$ in \eqnref{hamiltonian system 2}. 
 Let $g(x_i)\in\mathbb{R}^{2n}$ be the gradient of $H$ at a point $x_i$ in the phase space, which we know from data $(\dot{q},\dot{p})$. Then, 
\begin{equation}\begin{array}{rcl}
\frac{\partial}{\partial y}\hat H(y)\biggr\rvert_{y=x_i}&\approx& g(x_i),\\
\iff \frac{\partial}{\partial y}\hat k(y,X)^T\biggr\rvert_{y=x_i}k(X,X')^{-1}H(X)&\approx& g(x_i).\end{array}
\end{equation} 
Together with an arbitrary pinning term at a point $x_0$, the list of known derivatives leads to a linear system of $2nN+1$ equations, where we write
\begin{equation}
\frac{\partial}{\partial x}k(x_i,Y):=\frac{\partial}{\partial x}k(x,x')\biggr\rvert_{x=x_i,x'\in Y}
\end{equation}
for the derivative of the given kernel function $k$ with respect to its first argument, evaluated at a given point $x_i$ in the first argument and all points in the dataset $Y$ in the second argument. For each $x_i$, we thus have a (column) vector of $M$ derivative evaluations, which is transposed and stacked into a large matrix to form the full linear system
\begin{equation}\label{eq:GP linear system}
\underbrace{\left[\begin{matrix}
\frac{\partial}{\partial x}k(x_1,Y)^Tk(Y,Y')^{-1}\\\frac{\partial}{\partial x}k(x_2,Y)^Tk(Y,Y')^{-1}\\\dots\\\frac{\partial}{\partial x}k(x_N,Y)^Tk(Y,Y')^{-1}\\ k(x_0,Y)^Tk(Y,Y')^{-1}\
\end{matrix}\right]}_{\in\mathbb{R}^{(2nN+1)\times M}}\cdot \underbrace{\left[\begin{matrix}
H(Y)
\end{matrix}\right]}_{\in\mathbb{R}^{M}}=\underbrace{\left[\begin{matrix}g(X)\\H_0\end{matrix}\right]}_{\mathbb{R}^{2nN+1}}.
\end{equation}
We reiterate: $X$ is a data set of $N$ points where we know the derivatives of $H$ through $g(x_i)=(\frac{\partial H}{\partial q}(x_i), \frac{\partial H}{\partial p}(x_i))^T=(-\dot p_i,\dot q_i)^T\in\mathbb{R}^{2n}$. We evaluate on a fine grid $Y$ of $M$ points  (such that $k(Y,Y')\in\mathbb{R}^{M\times M}$, $\frac{\partial}{\partial x}k(x_i,Y)\in\mathbb{R}^{2n\times M}$) and have information $g(X)\in\mathbb{R}^{2nN}$ on a relatively small set of $N$ points called $X$ (black dots in
\figref{untransformed_gp}. The derivative of the Gaussian Process can be stated using the derivative of the kernel $k$ with respect to the first argument. The matrix inverse for $k(Y,Y')$ is approximated by the inverse of $(k(Y,Y')+\sigma^2 I)$, that is, through Tikhonov regularization with parameter $\sigma=10^{-5}$, as is the standard for Gaussian Process regression.
Solving this system of equations for $H(Y)$ yields the approximation for the solution to the PDE. \figref{untransformed_gp} shows the result, and table~I lists the mean squared error to the 625 training data points and an independently drawn set of 200 validation points in the same domain, where no derivatives were available.
See \cite{raissi_inferring_2017} for a more detailed discussion of the solution of PDE with Gaussian Processes.

\subsection{Approximation using an artificial neural network}\label{sec:nets for h}

Another possibility for learning the form of $H$ using data is to represent the function with an artificial neural network\cite{Goodfellow-et-al-2016} (ANN). We write
\begin{equation}
\label{eqn:mlp}
\mathbf x_l = \sigma_l(\mathbf x_{l-1} \cdot \mathbf W_l + \mathbf b_l), \quad l=1,\ldots,L+1
\end{equation}
where the activation function $\sigma_l$ is nonlinear (except where otherwise indicated, we used $\tanh$) for $l=1,\ldots,L$ (if $L\ge1$) and the identity for $l=L+1$.
The learnable parameters of this ANN are $\{(\mathbf W_l, \mathbf b_l)\}_{l=1,\ldots,L+1}$, and we gather all such learnable parameters from the multiple layers that may be used in one experiment into a parameter vector $w$.
If there are no hidden layers ($L=0$), then we learn an affine transformation
$\mathbf x_1=\mathbf x_0 \cdot \mathbf W + \mathbf b$.
This format provides a surrogate function $\hat H(q, p)=x_{L+1}$, where the input $x_{0}$ is the row vector $[q, p]$.
(Treating inputs as row vectors and using right-multiplication for weight matrices is convenient, 
as a whole batch of $N$ inputs can be presented as an $N$-by-$2$ array.)
For all the experiments shown here, this network
for computing $\hat H$
has two hidden layers of width 16.

Similarly, in the case(s) we need to learn additional transformations $\transLearned$ and $\transInvLearned$ (see \secref{transformed}), they are also learned using such networks.
%

We collect training data by sampling a number of initial conditions in the rectangle $(q,p)\in[-2\pi,2\pi]\times[-6,6]$, then simulate short trajectories from each to their final $(q,p)$ points. For each of these, we additionally evaluate $(\dot q, \dot p)$. Shuffling over simulations once per epoch, and dividing this dataset into batches, we then perform batchwise stochastic gradient descent to learn the parameters $w$ using an Adam optimizer on the objective function defined below.

For this paper, all neural networks were constructed and trained using TensorFlow, and the gradients necessary for evaluating the Hamiltonian loss terms in \eqnref{untransformedloss} were computed using TensorFlow's core automatic differentiation capabilities.

This objective function comprises a scalar function evaluated on each data 4-tuple $d = (q, p, \dot q, \dot p)$ in the batch, and then averaged over the batch. This scalar function is written as
\begin{equation}
\label{eqn:untransformedloss}
    f(q,p,\dot q, \dot p;\, w) = \sum_{k=1}^4 \weighting_k f_k,
\end{equation}
\begin{equation}
    \begin{array}{rclrcl}
    \\
    f_1 &=& \left( \Hder{p} - \dot q \right)^2,
    &
    f_2 &=& \left( \Hder{q} + \dot p \right)^2,
    \\
    f_3 &=& \left( \hat H(q_0, p_0) - H_0 \right)^2,
    &
    f_4 &=& \left(\Hder{q} \dot q + \Hder{p} \dot p \right)^2,
    \end{array}
\end{equation}
where the dependence on $w$ is through the learned Hamiltonian $\hat H$,
the loss-term weights $\weighting_k$ are chosen to emphasize certain properties of the problem thus posed,
and the partial derivatives of $\Hder{p}$ and $\Hder{q}$ are computed explicitly through automatic differentiation.
Except for $\weighting_2$, all $\weighting_k$ values are set to either $1$ or $0$ depending on whether the associated loss term is to be included or excluded.
Because of the square term in \eqnref{H true}, we set $\weighting_2$ arbitrarily to $10$ if nonzero, so the loss is not dominated by $f_1$. An alternative might be to set $\weighting_1$ to $1/10$.

\begin{figure}
\centering
\subfigure[][]{%
\label{fig:untransformed_truth}%
\includegraphics[width=\columnwidth]{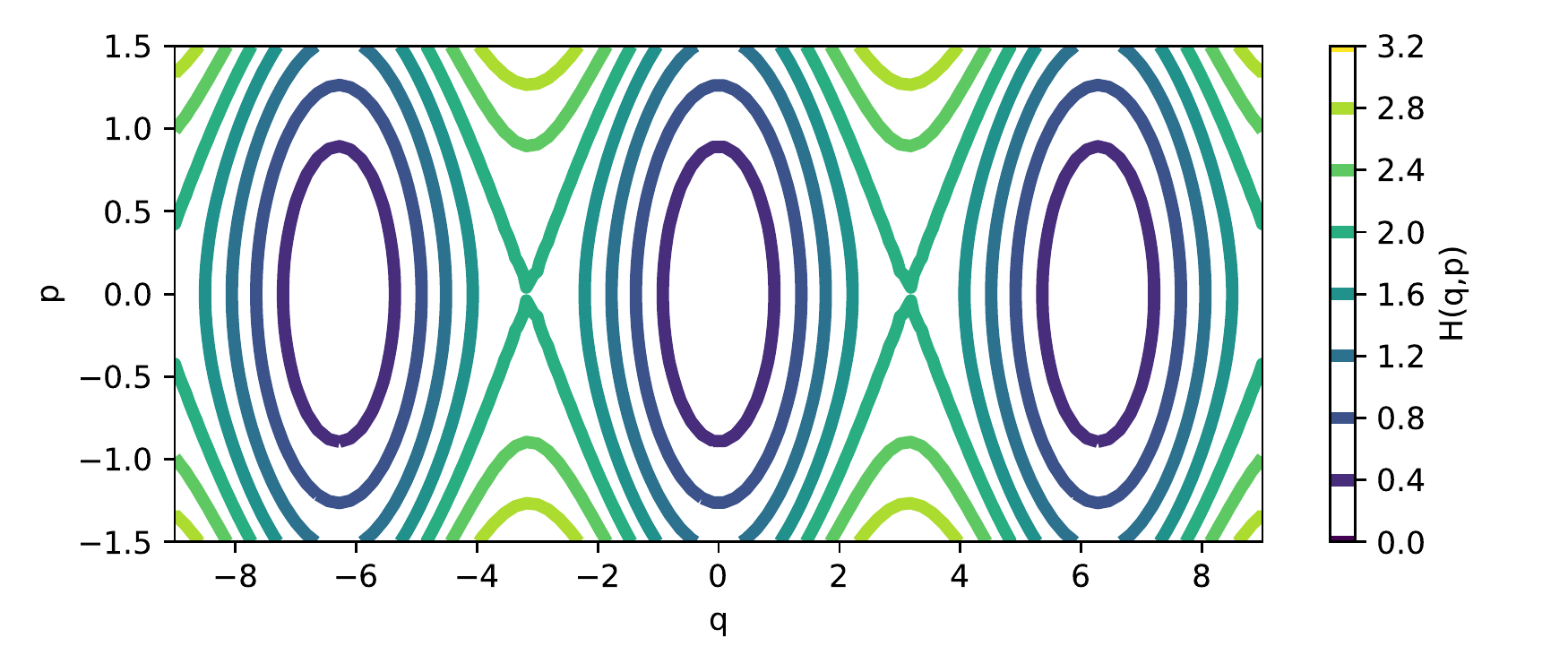}%
}

\subfigure[][]{
\label{fig:untransformed_gp}
\includegraphics[width=\columnwidth]{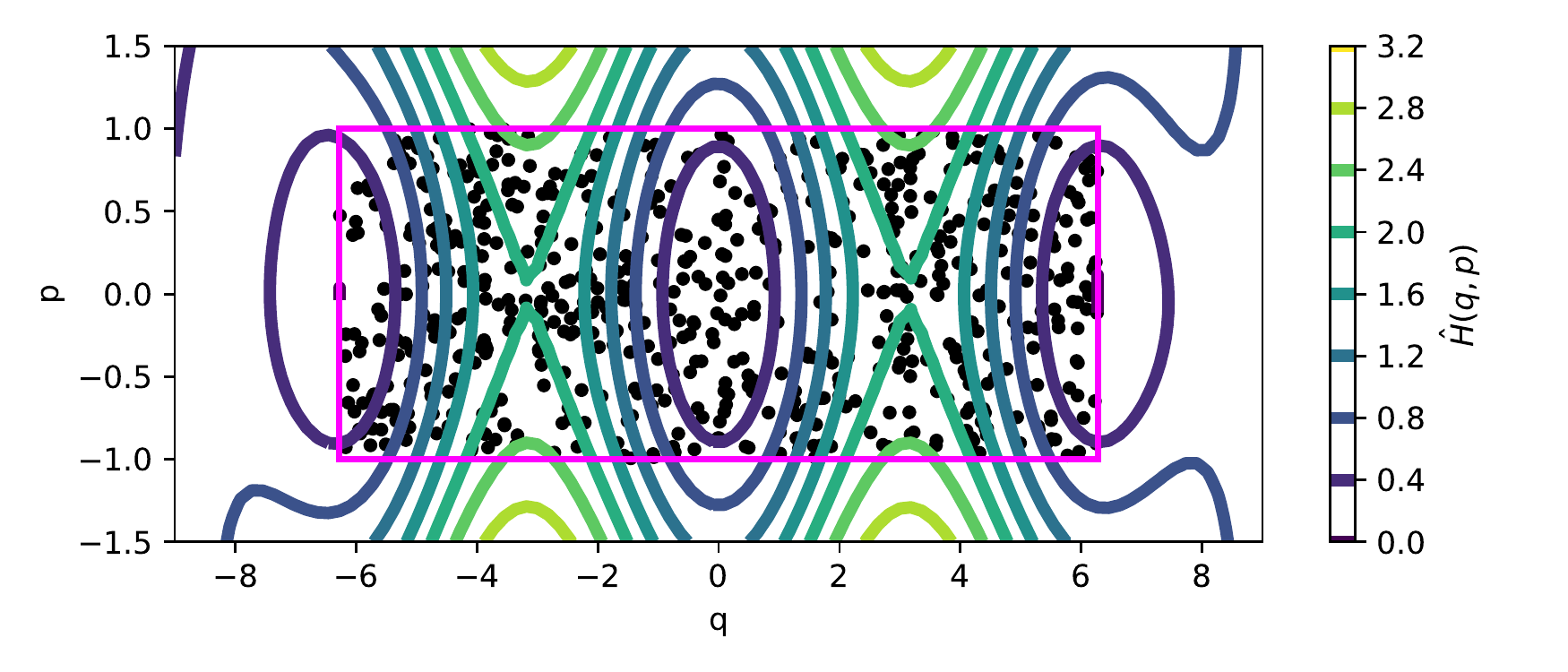}
}

\subfigure[][]{
\label{fig:untransformed_ann}
\includegraphics[width=\columnwidth]{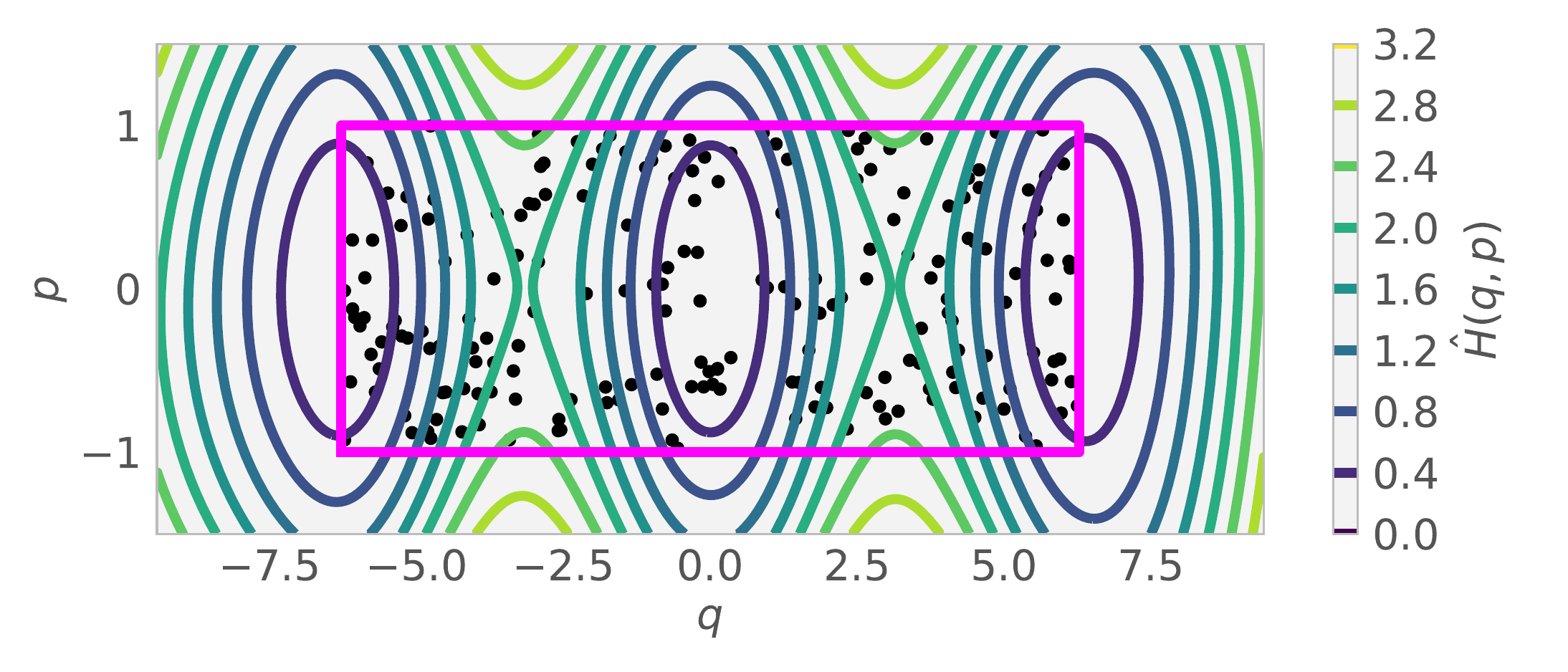}
}

\caption{
    \label{fig:gp_pendulum_hamiltonian_finegrid}
    \subref{fig:untransformed_truth}
    \textbf{Hamiltonian function of the pendulum system.}
    \subref{fig:untransformed_gp}
    \textbf{Approximation using Gaussian Processes}, solving equation \eqref{eq:GP linear system} on a fine grid $Y\subset[-9,9]\times[-1.5,1.5]$, with information $g$ about the derivative of $H$ on a set of $625$ randomly sampled points $X\subset[-2\pi,2\pi]\times[-1,1]$ (black points in the pink rectangle).
    \subref{fig:untransformed_ann}
    \textbf{Hamiltonian learned with a neural network}, with all loss terms included.
    Note that the learned function is only accurate where data sampling is sufficiently dense.
    For this experiment, we used the "softplus" activation function $\sigma_l(z) = \ln(e^z + 1)$. 
    The black points shown in panel (b) show all training data for the Gaussian Process. The points in panel (c) only show a subset of the training data used for the neural network (20000 in total).
    %
}
\end{figure}

Since equations (\ref{eqn:hamiltonian system 1}) and (\ref{eqn:hamiltonian system 2}) together imply (\ref{eqn:Hdot}),
any one of the three terms $f_1$, $f_2$, and $f_4$ could be dropped as redundant; 
therefore, we can set $\weighting_4$ to zero, 
but monitor $\dot{\hat{H}}$ as a useful sanity check on the accuracy of the learned solution.
In \figref{untransformed_ann}, we show the results of this process
with our default nonzero values for $\weighting_k$.

As an ablation study, we explored the effect (not shown here) of removing the first, second, and fourth terms.
By construction, the true $H_t(q, p)$ function is zero for all $(q,p)$. 
Note that this is only ever achieved to any degree in the central box in the figure, where data was densely sampled.
%
Removing $f_4$ made no visible difference in the quality of our $\hat H_t\approx 0$ approximation,
which was expected due to the redundancy in the set of 
equations (\ref{eqn:hamiltonian system 1}), (\ref{eqn:hamiltonian system 2}), and (\ref{eqn:Hdot}).
However, removing either $f_1$ or $f_2$ gives poor results across the figure,
despite the apparent redundancy of these terms with $f_4$.
This might be due to not balancing the contributions of the $\dot p$ and $\dot q$ terms, for which we attempted to compensate by unequal weighting values $\weighting_1$ and $\weighting_2$.
%

%

\section{Estimating Hamiltonian structure from observations}
\label{sec:transformed}

We now consider a set of observation functions $\trans : E \to \mathbb{R}^M$, $\transformLetter=(y_1,\dots,y_M)$, with $M\geq \dim E=2n$, such that $\trans$ is a diffeomorphism between the phase space $E$ and its image $\trans(E)$.
In this setting, the notion of a \textit{symplectomorphism} is important~\cite{almeida1992}. In general, a symplectomorphism is a diffeomorphism that leaves the symplectic structure on a manifold invariant. In our setting, a symplectomorphism of $E = Q \times P$ maps to a deformed space $\hat{E}=\hat{Q} \times \hat{P}$ where the system dynamics in the new variables $\hat{q}\in\hat{Q}$, $\hat{p}\in\hat{P}$ is again Hamiltonian, and conjugate to the original Hamiltonian dynamics. Not every diffeomorphism is a symplectomorphism, and we \textbf{do not assume} that $\trans$ is a symplectomorphism. A constant scaling of the coordinates is also not possible to distinguish from a scaling in the Hamiltonian function itself, so the recovered system will be a symplectomorphic copy of the original, scaled by an arbitrary constant.

\begin{figure}

\centering

\subfigure[][]{
    \label{fig:autoencoderpicture-diagram}
    \includegraphics[width=\subfigwidthfrac\textwidth]{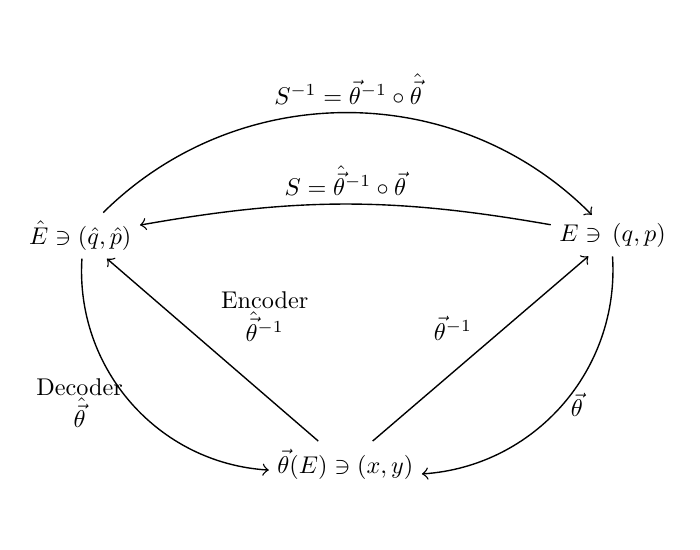}
}
    
\subfigure[][]{
    \label{fig:autoencoderpicture-transformation}
    \includegraphics[width=\subfigwidthfrac\textwidth]{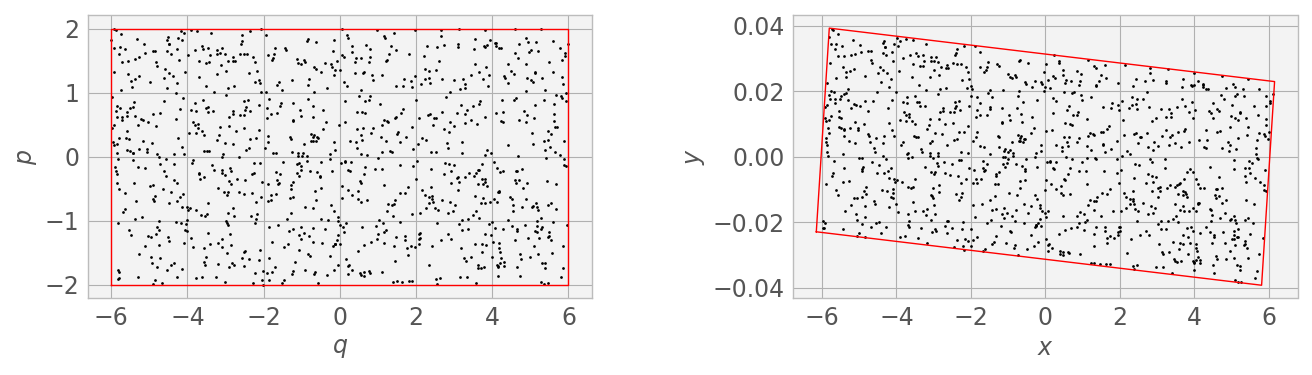}
}
    
\caption{
    \subref{fig:autoencoderpicture-diagram}
    \textbf{Illustration of the network structure.}
    Starting with observations of the (unknown) variables $(x,y)\in \trans(E)$, the autoencoder is structured to map via $\transInvLearned$ to the Hamiltonian form $(\hat q, \hat p)\in\hat E$, as well as map back to the observations through $\transLearned$. The complete process involves estimating a Hamiltonian for (the symplectomorphic copies of) $p,q$ during training of the autoencoder.
    \subref{fig:autoencoderpicture-transformation}
    \textbf{The transformation $\trans$ from the original space $q,p$ to the observed space $x,y$, for the linear example of \secref{linearTransformation}.
    }
}
\end{figure}

In the setting of this section, we do not assume access to $E$, $H$, or the explicit form of $\trans$. Only a collection of points $\trans_i$ and` time derivatives $\frac{d}{dt}\trans_i$ in the image $\trans(E)$ is available. We describe an approach to approximate a new map $\transInvLearned:\trans(E)\to\hat{E}$ into a symplectomorphic copy of $E$ through an autoencoder ~\cite{Goodfellow-et-al-2016}, such that the transformed system in $\hat{E}$ is conjugate to the original Hamiltonian system in $E$. 
Upon convergence, and if we had access to $\trans$, the map $\transInvLearned \circ \trans \equiv \symplecto : E \to \hat E$ would approximate a symplectomorphism,
and $\transInv \circ \transLearned \equiv \symplecto^{-1}$ would be its inverse.
During the estimation of $\transInvLearned$, we simultaneously approximate the new Hamiltonian function $\hat{H} : \hat{E} \to \mathbb{R}$.
\figref{autoencoderpicture-diagram} visualizes the general approach, where only the information $(x,y)_i$ and $\frac{d}{dt}(x,y)_i$ is available to the procedure, while $\transInvLearned$ and a Hamiltonian $\hat H$ on $\hat E$ are constructed numerically.

An interesting and important feature common to the next three examples 
is that one cannot expect to systematically recover the original $(q,p)$ values from the given observation data $(x,y)$. Only symplectic transformations $(\hat q,\hat p)$ can be recovered, which are enough to define a Hamiltonian. Once the coordinates $(\hat q,\hat p)$ are fixed, the Hamiltonian function \textit{in these coordinates} is unique up to an additive constant.

Table~I contains training and validation loss of the networks for all experiments.
Additionally, we trained the network from \secref{images} with only 331 images, and did observe a significantly higher validation loss (not shown), consistent with overfitting.
\begin{table*}
    \centering
    \label{tab:cant_overfit}
    \begin{tabular}{cccccc}
         & \# Parameters & \# Training Points & \# Validation Points & Training Loss & Validation Loss \\
        \hline
        \secref{gps for h} & non-parametric & 625 & 200 &
        $2.2\cdot10^{-5}$ &
        $3.5\cdot10^{-5}$ \\
        \secref{nets for h} & 337 & 20000 & 200 &
        $6.8\cdot10^{-4}$ &
        $7.4\cdot10^{-4}$ \\
        \secref{linearTransformation} & 345 & 20000 & 1000 &
        $5.5\cdot10^{-6}$ &
        $5.0\cdot10^{-6}$ \\
        \secref{nonlinearTransformation} & 511 & 19200 & 200 &
        $1.4\cdot10^{-4}$ &
        $1.8\cdot10^{-4}$ \\
        \secref{images} & 755 & 17489 & 200 
        & 0.51 & 0.54
        \\ \hline &&&&&
    \end{tabular}
    \caption{
    \textbf{Comparison of loss between training and held-out validation data.}
    As the sizes of our generated datasets are always much larger
    than the numbers of trainable parameters, overfitting is not an issue, and the validation loss is always
    indistinguishable from the training loss.
The first row includes the results from the Gaussian process approximation (a non-parametric method) using all training data for inference. Only the hyper-parameters $\epsilon$ and $\sigma$ needed to be adjusted.
    For the first row, the loss reported is the MSE of the Gaussian Process regression. For other rows, the loss is computed by either \eqnref{untransformedloss} or \eqnref{transformedloss} as appropriate.}
\end{table*}


\subsection{A composite loss function for the joint learning of a transformation and a Hamiltonian}
\label{sec:transformationLoss}
The following loss function is used to train an autoencoder component together with a Hamiltonian function approximation network component:
%
%
\begin{equation}
    \begin{array}{rcl}
    \\
    f(\hat q, \hat p,\dot{\hat q}, \dot{\hat p};\,w) 
&=& \sum_{k} \weighting_k f_k,
\label{eqn:transformedloss}
    \\
    f_1 &=& \left( \Hder{p} - \dot{\hat{q}} \right)^2,
    \\
    f_2 &=& \left( \Hder{q} + \dot{\hat{p}} \right)^2,
    \\
    f_3 &=& \left( \hat H(\transInvLearned(x_0, y_0)) - H_0 \right)^2,\\
    f_4 &=& \left(\dot{\hat{H}}\right)^2 = \left(\Hder{\hat q} \dot{\hat q} + \Hder{\hat p} \dot{\hat p} \right)^2,
    \\
    f_5 &=& ||\transLearned(\transInvLearned(x,y)) - [x,y] ||^2,\\
    f_6 &=& \left(
    \det(D\transInvLearned)
\right)^{-2},
    \end{array}
\end{equation}
where
the dependence on $w$ is through the learned Hamiltonian $\hat H$
and the learned transformations $\transLearned$ and $\transInvLearned$,
and
the time derivatives in the space $\hat E$ are computed as 
$\dot{\hat{q}}=\dot x \pder{\hat q}{x} + \dot y \pder{\hat q}{y}$
and 
$\dot{\hat{p}}=\dot x \pder{\hat p}{x} + \dot y \pder{\hat p}{y}$,
using the Jacobian 
$D\transInvLearned=\left[
\begin{array}{cc}
    \pder{\hat q}{x} & \pder{\hat q}{y} \\
    \pder{\hat p}{x} & \pder{\hat p}{y} \\
\end{array}
\right]$
of the transformation $\transInvLearned$
(computed pointwise with automatic differentiation).
When we learn an (especially nonlinear) transformation $\transInvLearned$,
in addition to the Hamiltonian $\hat H(\hat q, \hat p)$,
including the $f_4$ term in the composite loss
can have a detrimental effect on the learned transformation.
There exists an easily-encountered naive local minimum in which $\transInvLearned$
maps all of the sampled values from $\trans(E)\ni(x,y)$
to a single point in $\hat E$,
and the Hamiltonian learned is merely the constant function
at the pinning value, $\hat H(\hat q, \hat p) = \hat H_0$.
In this state
(or an approximation of this state), all of 
$\pder{H}{\hat q}$,
$\pder{H}{\hat p}$,
and the elements of $D\transInvLearned$
are zero, so the loss 
(with terms $f_1$, $f_2$, $f_3$, and optionally $f_4$)
is zero (resp., small).
A related failure is that in which the input in $\trans(E)$ is collapsed by $\transInvLearned$ to a line or curve in $\hat E$.

To alleviate both of these problems we 
added a new loss component $f_6$.
That is, we require that the learned transformation not collapse the input.
It is sufficient for the corresponding weighting factor $\weighting_6$ to be a very small nonzero value (e.g. $10^{-6}$).
The addition of $f_6$ to our loss helps us to avoid falling early in training into the unrecoverable local minimum described above,
and also helps keep the scale of the transformed variables $\hat q$ and $\hat p$ macroscopic.


\subsection{Example: linear transformation of the pendulum}
\label{sec:linearTransformation}
\newcommand{\rotationangle}{\rho}

We generate data from a rectangular region $x\in[-1,1]$, $y\in[-1,1]$, then transform the region linearly with $\transInv(x,y)=A^{-1}[x,y]^T=[q,p]^T$. The matrix $A^{-1}$ is the inverse of $A = R \cdot \Lambda$; a scaling followed by a rotation where $\Lambda=\left[\begin{array}{cc}\lambda_1 & 0\\ 0 &\lambda_2\end{array}\right]$, $\lambda_1=1$,  $\lambda_2=64$,  $R=\left[\begin{array}{cc}\cos\rotationangle & -\sin\rotationangle\\ \sin\rotationangle & \cos\rotationangle \end{array}\right]$, and $\rotationangle=5^\circ$.
Our observation data $\trans(E)$ is thus given by $[x, y]^T = A \cdot [q, p]^T$. 
Using the true Hamiltonian $H(q,p)=p^2/2 + (1-\cos q)$, we additionally compute true values for $dq/dt$ and $dp/dt$, and then use $A$ to propagate these to $x$ and $y$ via
$\frac{\diff x}{\diff t}=\frac{\partial x}{\partial q} \frac{\diff q}{\diff t} + \frac{\partial y}{\partial p}\frac{\diff p}{\diff t}$
and similar for $y$,
where the partial derivatives are computed analytically (here, just the elements of $A$ itself).

Our network is then presented with observation data $x, y$ and its corresponding time-derivatives. Its task is to learn $\hat A$ and $\hat A^{-1}$, which convert to and from variables $\hat q, \hat p$ (symplectomorphic to the original $q,p$); and a Hamiltonian $\hat H$ in this new space. 
When evaluating the loss, the time derivatives of $\hat q$ and $\hat p$ are likewise computed via automatic differentiation using the chain rule through the learned transformation, e.g. as $\frac{\diff \hat q}{\diff t}=\frac{\partial \hat q}{\partial x} \frac{\diff x}{\diff t} + \frac{\partial \hat q}{\partial y}\frac{\diff y}{\diff t}$.
Note also that $[\hat q, \hat p]^T =\hat A^{-1} \cdot A \cdot [q, p]^T$, so if the original space $E$ could be found, $\hat A$ would satisfy $\hat A \cdot A^{-1}=I$. This cannot be expected given only the data in $\trans(E)$; we can only be sure that $\hat A \cdot A^{-1}$ approximates a symplectomorphism of the original $E$.

\begin{figure}
\centering

\subfigure[]{
    \label{fig:linear_results-H}
    \includegraphics[width=1\columnwidth]{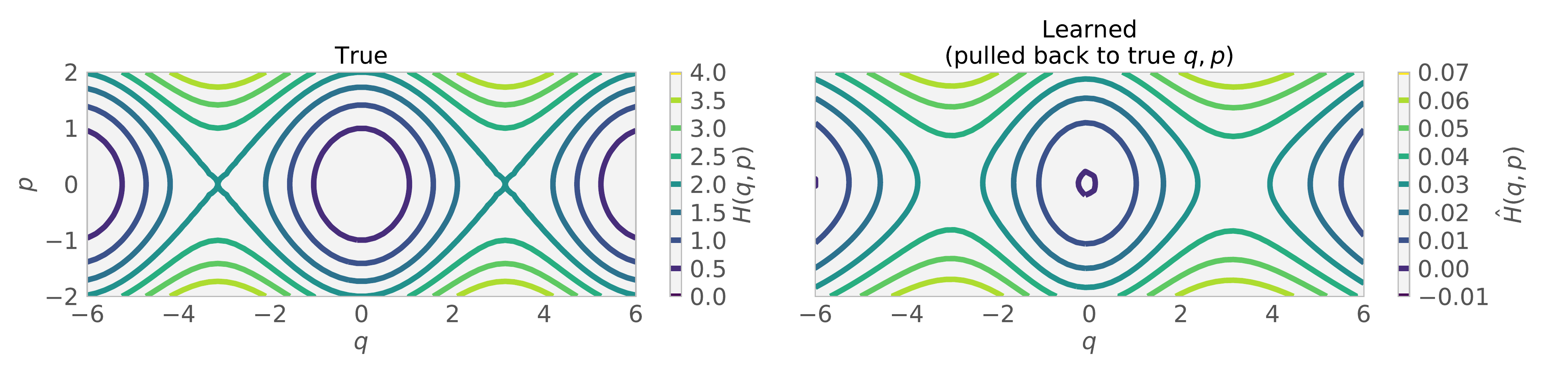}
}

\subfigure[]{
    \label{fig:linear_results-transformation}
    \includegraphics[width=1\columnwidth]{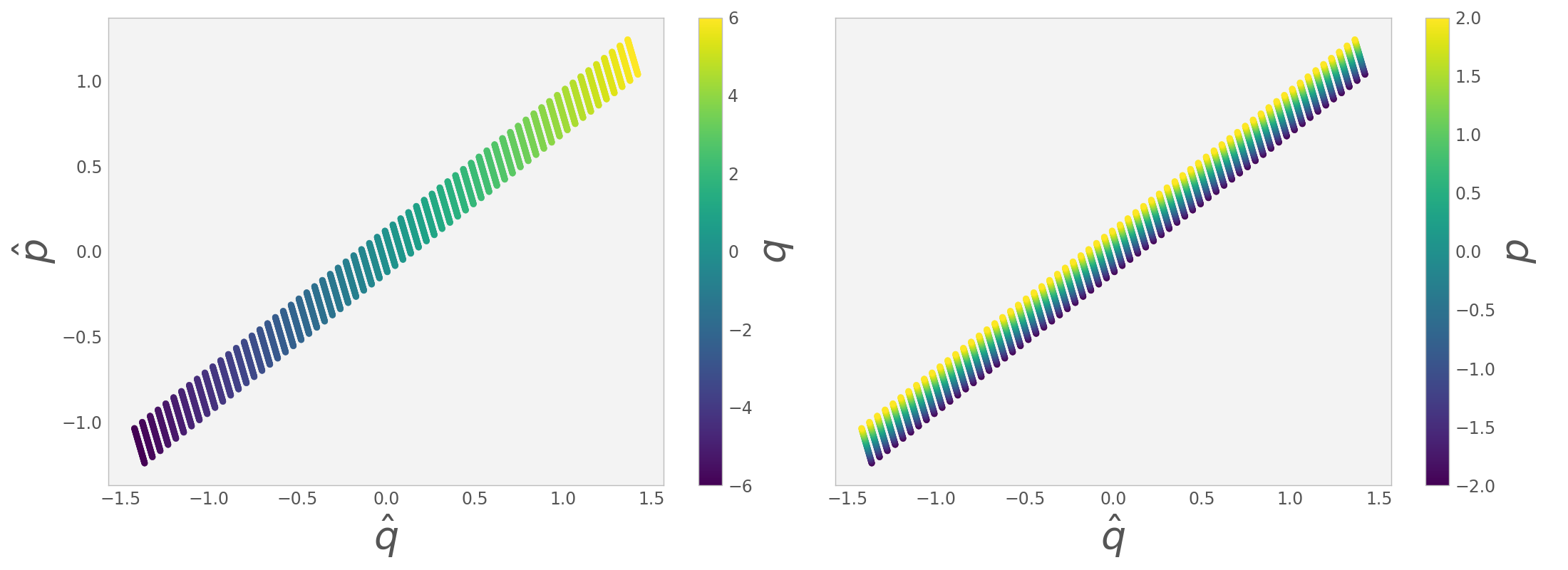}
}

\caption{\label{fig:linear_results}
    \textbf{Learned $\hat H(\hat q, \hat p)$ after a linear transformation $\trans$.}
    \subref{fig:linear_results-H} (Left)
        \textbf{The true function} $H(q,p)$ evaluated on a grid of $q,p$ values.
    \subref{fig:linear_results-H} (Right)
        \textbf{The learned function} on $\hat q, \hat p$, 
        pulled back onto the original $q,p$ space.
    %
    %
    \subref{fig:linear_results-transformation}
    \textbf{Linear symplectomorphism} $\symplecto=\transInvLearned \circ \trans$ between the original and the learned space.
    %
    %
}
\end{figure}

We could learn $\transInvLearned$ from a general class of nonlinear functions, as a small $\tanh$ neural network, but here we simply learn $\hat A$ and $\hat A^{-1}$ as linear transformations (that is, we have a linear ``neural network'', where $L=0$ in \eqnref{mlp}, and $\mathbf b_1$ is fixed as $\mathbf 0$). As we include the reconstruction error of this autoencoder in our loss function, $\hat A^{-1}$ is constrained to be the inverse of $\hat A$ to a precision no worse than the $f_1$ and $f_2$ terms in \eqnref{transformedloss}, after all three are scaled by their corresponding $\weighting_k$ values.
In fact, for the linear case, initially the autoencoder's contribution to the loss is significantly lower than the Hamiltonian components (see \figref{linear_results}), but, as training proceeds and the $f_1$ and $f_2$ terms are improved (at the initial expense of raising the autoencoder loss), larger reductions in loss are possible by optimizing $\hat H$ rather than $\transLearned$, so $f_5$ is decreased as quickly as (the larger of) $f_1$ or $f_2$.
That is, the autoencoder portion of the loss falls quickly to the level where it no longer contributes to the total loss given its weighting in the loss sum.

We find that the learned symplectomorphism $\symplecto(q,p)=\hat{A}^{-1} \cdot A \cdot [q, p]^T$,
depicted in its $q$ portion in \figref{linear_results},
preserves $q$ unmixed with $p$ in one or the other of its two discovered coordinates.
This is because both (a) $(q,p)\mapsto (p,-q)$ as well as (b) $(q,p)\mapsto(q,p+f(q))$ for any smooth function $f$ are symplectomorphisms. They are special because $H(q,p)=\hat H(\hat q(q,p), \hat p(q,p))$, i.e. they even preserve the Hamiltonian formulation.
For the map (a), the transformation of the Hamiltonian can be seen from the following derivation.
\begin{eqnarray}
\dot {\hat q} &=&\dot p=-\partial H/\partial q=\partial \hat H/\partial \hat p,\label{eqn:htransform 1}\\
\dot {\hat p}&=&-\dot q=-\partial H/\partial p=-\partial \hat H/\partial \hat q,\label{eqn:htransform 2}\\
\partial \hat H/\partial q&=&\partial\hat H/\partial \hat q \cdot \underbrace{\partial \hat q/\partial q}_{=0} + \partial\hat H/\partial \hat p \cdot \underbrace{\partial \hat p/\partial q}_{=-1} =\\&=&-\partial \hat H/\partial\hat p=\partial H/\partial q\label{eqn:htransform 3}\nonumber\\
\partial \hat H/\partial p&=&\partial\hat H/\partial \hat q \cdot \underbrace{\partial \hat q/\partial p}_{=1} + \partial\hat H/\partial \hat p \cdot \underbrace{\partial \hat p/\partial p}_{=0} =\\&=&\partial \hat H/\partial\hat q=\partial H/\partial p.\nonumber\label{eqn:htransform 4}
\end{eqnarray}
Here, the first equality of (\ref{eqn:htransform 1}, \ref{eqn:htransform 2}) follows from the map and the last equality of (\ref{eqn:htransform 1}, \ref{eqn:htransform 2}) follows from the requirement that $\hat q,\hat p$ follow Hamiltonian dynamics with respect to the new Hamiltonian $\hat H$.
Equations (\ref{eqn:htransform 3},\ref{eqn:htransform 4}) then show that the new Hamiltonian is the same as the old one (modulo an additive constant) when considered as a map on the old coordinates. 
%
%

\subsection{Example: nonlinear transformation of the pendulum}
\label{sec:nonlinearTransformation}

In addition to the linear $\trans$ of \secref{linearTransformation},
we show comparable results for a nonlinear transformation $\trans$ and learned $\transLearned$.
Specifically, we transform the data through $(x,y) = \trans(q,p)$ where
\begin{equation}
\label{eqn:nonlinear_trans}
\begin{array}{rclrcl}
 a &=& q / 20 
 &
 b &=& p / 10 \\
 x &=& a + (b + a^2)^2 
 &
 y &=& b + a^2,\\
 \end{array}
\end{equation}
the inverse of which is given by
$q = (x - y^2) 20$
and
$p = (y - x^2 + 2 x y^2 - y^4) 10$.
%
We use the analytical Jacobian of this $\trans$
to compute the necessary $\dot x$ and $\dot y$ for input to our network.

We proceed as before, except that we no longer restrict the form of the learned $\transLearned$ and $\transInvLearned$ to linear transformations, but instead allow small multi-layer perceptrons of a form similar to that used for $\hat H$.

\begin{figure}
\centering

\subfigure[]{
    \label{fig:learned_nonlinear-H}
    \includegraphics[width=\columnwidth]{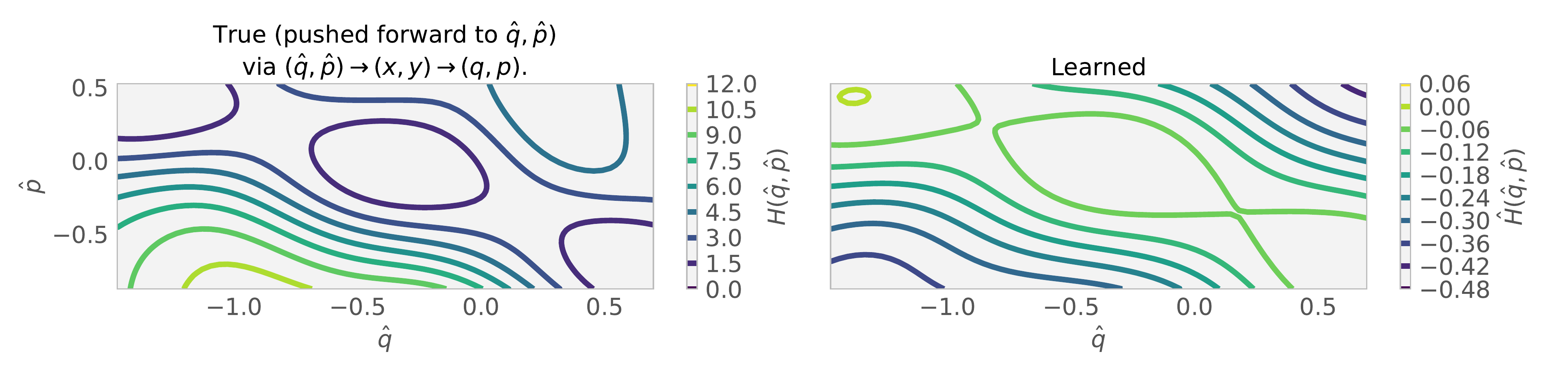}
}

\subfigure[]{
    \label{fig:learned_nonlinear-transformation}
    \includegraphics[width=\columnwidth]{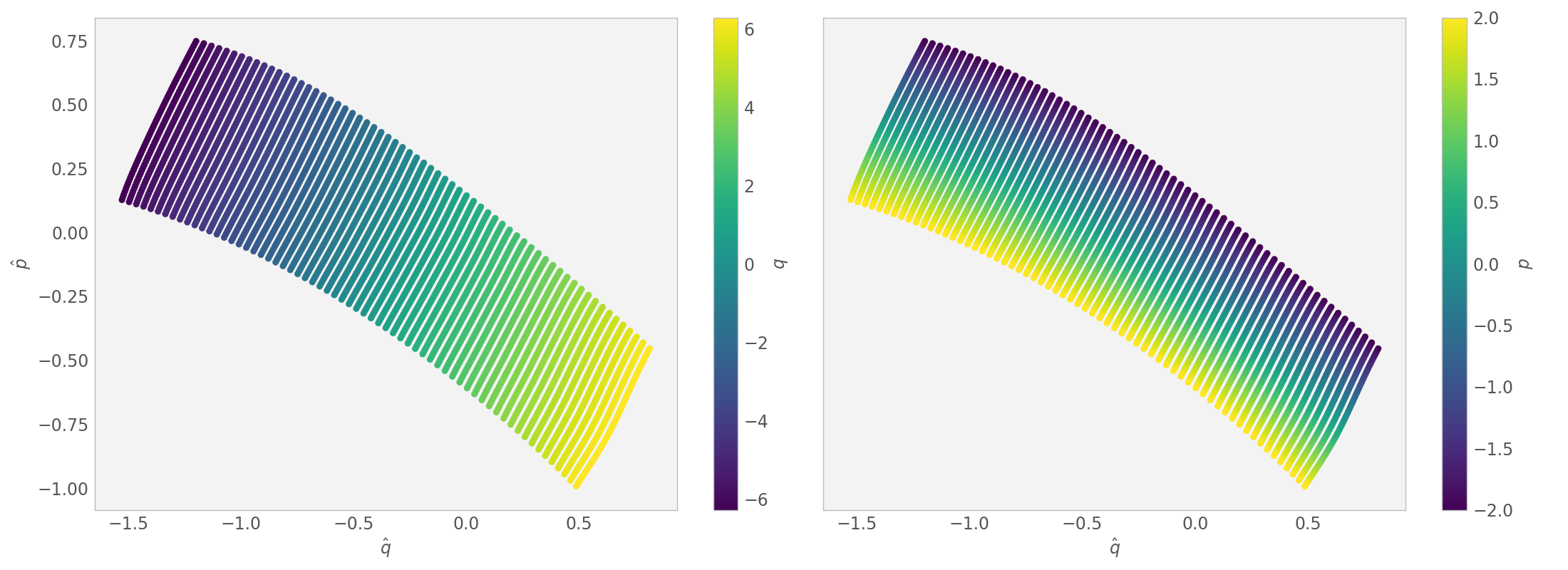}
}

\caption{
    \label{fig:learned_nonlinear}
    \textbf{Results after a nonlinear transformation $\trans$.}
    %
    %
    \subref{fig:learned_nonlinear-H} (Left)
    \textbf{
        The true  function $H(q,p)$, pushed forward to $\hat q, \hat p$.
    }
    \subref{fig:learned_nonlinear-transformation}
    \textbf{
    The learned function on $\hat q$, $\hat p$.}
    A grid is now taken in $\hat E$, and transformed through $\symplecto^{-1}$
    for plotting $H$ at corresponding $q,p$ points.
    As is also true for the linear case, the sign of the learned  $\hat H$
    may be flipped depending on whether we learn a representation of $q$ via $\hat q$ or $\hat p$
    which is monotonically increasing or decreasing.
    \textbf{Bottom: Nonlinear symplectomorphism $\symplecto = \transInvLearned \circ \trans$ between the original and the learned space.}
    Again, we find that $q$ is preserved (nearly) unmixed in the discovered space $\hat E \ni (\hat q, \hat p)$.
        For this experiment, the encoder $\transInvLearned$ and decoder $\transLearned$ each have three hidden layers of width five.
}
\end{figure}

The resulting induced symplectomorphism is again
one which appears to preserve an approximately monotonically increasing or decreasing $q$ in either $\hat q$ or $\hat p$.
%
This can be seen in \figref{learned_nonlinear}.

\subsection{Example: constructing a Hamiltonian system from nonlinear, high-dimensional observations of $q,p$}
\label{sec:images}
\begin{figure}
\centering

\subfigure[]{
    \label{fig:movie-images}
    \includegraphics[width=0.3\columnwidth]{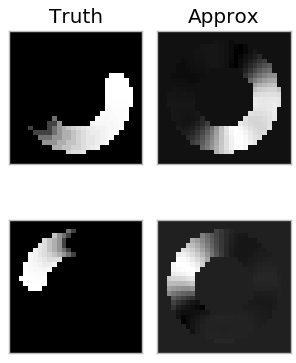}
}
\subfigure[]{
    \label{fig:movie-H}
    \includegraphics[width=0.6\columnwidth]{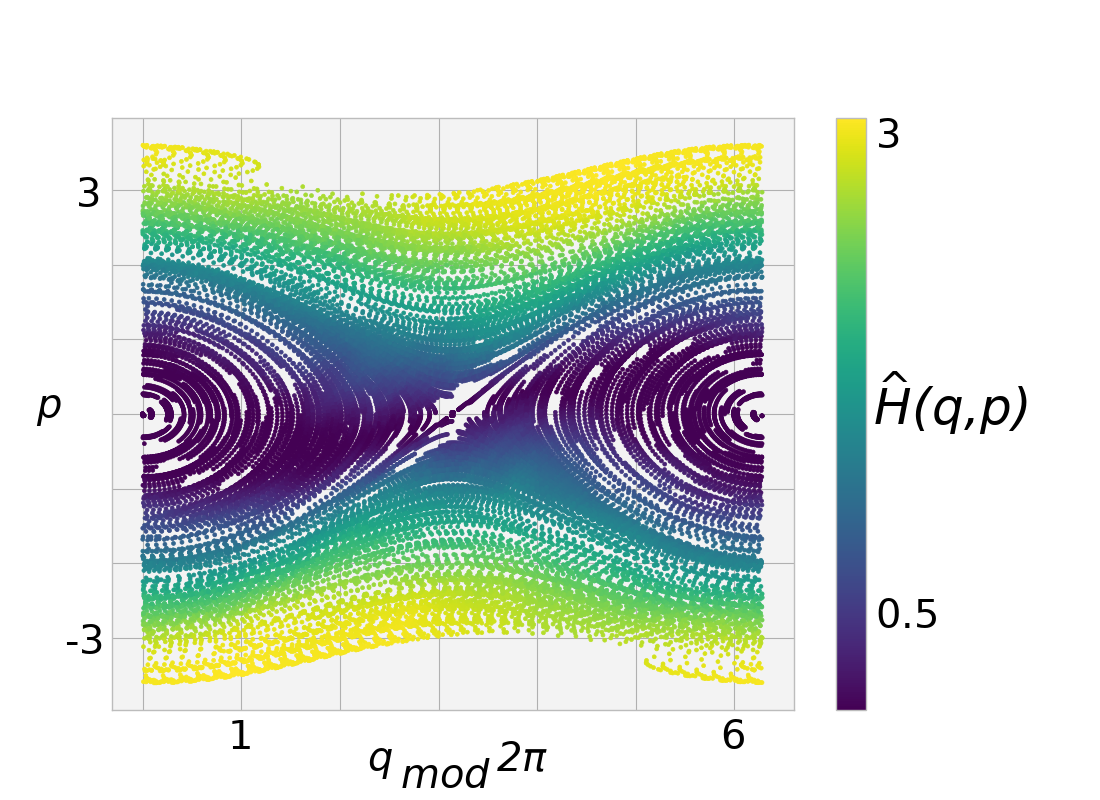}
}

\caption{
    \label{fig:movie}
    \subref{fig:movie-images}
    \textbf{PCA autoencoder reconstructions.}
    The autoencoder is trained to reproduce PCA projections of the monochrome images on the left,
    where velocity $p$ information is available in the length of the trail dragged behind the moving pendulum head.
    We show in the column ``Approx'' these reconstructions passed back through the approximate inverse of the PCA projection.
    Note that only the magnitude of $p$ is preserved,
    which can be predicted by the fact that only the square of $p$ is required for computing $H$.
    \subref{fig:movie-H}
    \textbf{Learned Hamiltonian.}
    For each of the images $\imagex$ in our dataset, we have an associated known $q,p$ pair. Here, we plot these values, colored by the learned $\hat H(\imagex)$.
        For this experiment, the encoder $\transInvLearned$ has two hidden layers of widths six and four respectively,
        and the decoder $\transLearned$ mirrors this structure.
}
\end{figure}

As a further demonstration of the method, we use a graphical rendering of the moving pendulum example from before as the transformation $\trans$ from the intrinsic state $(q,p)$ to an image $\imagex$ as our high-dimensional observable.
We use a symplectic semi-implicit Euler's method
\begin{equation}
\label{eqn:symplecticEuler}
    \begin{array}{rcl}
    p(\tau) &=& 
        p(0) + \tau \cdot \left.\dot p    \right|_{q(0), p(0)} \\
    q(\tau) &=&
        q(0) + \tau \cdot \left. \dot q\right|_{q(0), p(\tau)}
    \end{array}
\end{equation}
to generate $q(t_i),p(t_i)$ trajectories for various initial conditions,
and then a simple graphical renderer to display these as images (see \figref{movie-images}).
When rendering our video frames, we drag a tail of decaying images behind the moving pendulum head, so that information about both position $q$ and velocity $p$ is present in each rendered frame.
    This is done by
    iterating over each $q(t_i),p(t_i)$ trajectory,
    and, for each $t_i$,
    (1) multiplying the entire current image by a cooling factor of $\sim 0.96$,
    (2) adding a constant heating amount to the image in a circle of fixed radius centered around the current $\cos(q), \sin(q)$ point, 
    and
    (3) clipping the image per-pixel to lie within $[0, 1]$.
    Samples of the resulting images
    are visible in \figref{movie-images}.

    Though we do not use $p(t_i)$ directly in this procedure, is value is observed in the length of the resulting tail dragged behind the moving pendulum head.
    We create trajectories long enough that the effect of the initial formation of the tail 
    (during which its length is not necessarily a good indicator of $p$) 
    is not visible any more,
    and then use only the final two observations from these trajectories.

%
In order to make the approach agnostic to the data, we do not want to assume that the space $\hat E$ is periodic, so instead, we use a four-dimensional phase space with elements $\hatzvec=[\hat q_1,\hat q_2, \hat p_1, \hat p_2]=[\hatqvec, \hatpvec]$ and consider the splitting into $(\hat q_1, \hat q_2)$ and $(\hat p_1, \hat p_2)$ during training.
In the space of input images, the manifold does not fill up four-dimensional space, but a cylinder, which is mapped to the four-dimensional encoding layer by the autoencoder. 


%
In addition, to simplify the learning problem, we learn $\transInvLearned$ as the combination of a projection onto the first twenty principal components of the training dataset
followed by a dense autoencoder,
reserving learning $\transInvLearned$ as an end-to-end convolutional autoencoder for future work.
The encoding provides $\hatzvec$ and, as before, we learn $\transInvLearned$ in tandem with $\hat H$,
where now the conditions of \eqnref{transformedloss} are upgraded to vector equivalents to accommodate $\hatzvec$.
    

In \secref{transformationLoss}, we added a loss term proportional to the reciprocal of the determinant of the transformation's Jacobian in order to avoid transformations that collapsed the phase space.
Here, this was not such an issue--of course, some collapsing of the high-dimensional representation is obviously required.
Instead, a common mode of failure turned out to be learning constant $\hat H$ functions, which automatically satisfy the Hamiltonian requirements (a constant is naturally a conserved quantity).
To avoid this, we considered several possible ways to promote a non-flat $\hat H$ function,
ultimately settling on (a) adding a term that encouraged the standard deviation of $\hat H$ values to be nonzero, and (b) minimizing not just the mean squared error in our $f_1$ and $f_2$ terms, but also the max squared error, to avoid trivial or bi-level $\hat H(q, p)$  functions.
%
The spherical Gaussian prior
used in training variational autoencoders 
\cite{toth_hamiltonian_2019}
also avoids, as a byproduct, learning constant functions.

The result, shown in \figref{movie-H}, was a pulled-back $\hat H(q,p)$ function that at least in broad strokes resembles the truth,
and
that satisfies $\diff \hat H/\diff t\approx 0$ (typically about $10^{-2}$).

\section{Conclusions}
We described an approach to approximate Hamiltonian systems from observation data. 
It is a completely data-driven pipeline to (a) construct an appropriate phase space and (b) approximate a Hamiltonian function on the new phase space, from nonlinear, possibly high-dimensional observations (here, movies).

When only transformations of the original Hamiltonian phase space can be observed, 
 it is only possible to recover a symplectic copy of the original phase space, with additional freedom in a constant scaling of the coordinates, and an additive constant to the Hamiltonian function.
If no additional information about the original space is available, this is a fundamental limitation, and not particular to our approach. It is not necessary that there is an ``original phase space'' at all, and so the resulting symplectic phase space is by no means unique. The choice of a single such space has to rely on other factors, such as, possibly, interpretability by humans, or simplicity of the equations.

The approach may be extended to time-dependent Hamiltonian functions. This would allow us to cope with certain dissipative systems\cite{mcdonald_hamiltonian_nodate}. An even broader extension may allow transformations to arbitrary normal forms as the ``target vector field'', and thus would not be constrained to Hamiltonian systems. In the general case, it will become important to explore whether the transformation we approximate remains bounded over our data, or whether it starts showing signs of approaching a singularity, suggesting that the problem may not be solvable.



\begin{acknowledgements}
    This work was funded by the
    US Army Research Office (ARO) through a
    Multidisciplinary University Research Initiative (MURI) 
    and 
    by the Defense Advanced Research Projects Agency (DARPA) through their Physics of Artificial Intelligence (PAI) program.
\end{acknowledgements}

\section*{Bibliography}
\bibliographystyle{plain}
\bibliography{hamiltonian_literature}

\end{document}